\newcommand{\LiCuO}{LiCu$_2$O$_2$}

\documentclass[aps,prl,twocolumn,superscriptaddress]{revtex4}
\usepackage{graphicx}
\usepackage{bm}

\begin{document}

\title{Helimagnetism in a disordered and geometrically frustrated quantum spin chain.}

\author{T. Masuda}
\affiliation{Condensed Matter Sciences Division, Oak Ridge
National Laboratory, Oak Ridge, TN 37831-6393, USA.}
\author{A. Zheludev}
\affiliation{Condensed Matter Sciences Division, Oak Ridge
National Laboratory, Oak Ridge, TN 37831-6393, USA.}
\author{A. Bush}
\affiliation{Moscow Institute of Radiotechnics, Electronics and
Automation, Moscow 117464, Russia.}
\author{M. Markina}
\affiliation{Low Temperature Physics Department, Moscow State
University, Moscow 119992, Russia.}

\author{ A. Vasiliev}
\affiliation{Low Temperature Physics Department, Moscow State
University, Moscow 119992, Russia.}

\date{\today}
\begin{abstract}
Neutron diffraction and bulk measurements are used to determine
the nature of the low-temperature ordered state in a $S=1/2$
spin-chain compound with competing interactions. The magnetic
structure is found to be helimagnetic, with a propagation vector
$(0.5,\zeta,0)$, $\zeta=0.174$. The nearest-neighbor exchange
constant and frustration ratio are estimated to be $J=5.8$~meV and
$J_2/J_1=0.29$, respectively. For idealized spin chains these
parameter values would signify a gapped spin-liquid ground state.
Long-range ordering is attributed to intrinsic non-stoichiometry.
\end{abstract}


\maketitle \narrowtext

In recent years the focus in quantum and low-dimensional magnetism
has shifted from the simplest nearest-neighbor chain models to
more complex spin networks, effects of impurities, disorder, and
inter-chain coupling. Of particular interest are cases that
involve strong geometric frustration of interactions. Any of these
effects may either enhance or suppress quantum fluctuations, and
thus play a crucial role in determining the ground state and
magnetic properties. Even the simplest frustrated models, such as
the $S=1/2$ spin chain with competing nearest-neighbor (nn)
next-nearest-neighbor (nnn) antiferromagnetic (AF) interactions,
have a rich phase diagram that includes both gapped and
quantum-critical gapless phases \cite{nnn}. Among the latter is,
for example, a unique disordered chiral state \cite{Nersesyan98},
which is all that quantum fluctuations preserve of helimagnetism,
realized in the classical version of the model. The challenge is
to find experimental realizations of frustrated-chain constructs,
and to understand the effect of disorder,  impurities and extended
interactions in such materials.

The charge-ordered compound \LiCuO
\cite{Vorotynov,Fritschij98,Roessli,Zvyagin,Zatsepin}, may, in
fact, be just the right model system, featuring $S=1/2$ chains
with strong nnn interactions. Paradoxically, \LiCuO\ possesses
characteristics of a quantum-disordered ``dimer liquid'' state,
including a gap $\Delta\approx 6$~meV in the magnetic excitation
spectrum \cite{Zvyagin}, and yet undergoes at least one magnetic
ordering transition, at $T_c\approx 23$~K
\cite{Vorotynov,Roessli,Zvyagin}. The phenomenon is poorly
understood, and the key missing pieces of information are the
actual structure of the ordered phase and the magnitude of
frustration. These issues are topics of an ongoing controversy.
$\mu$-SR studies indicate a single complex non-collinear state
below $T_c\approx 22.5$~K, with a precursor transition at
$T_1=24$~K. Some recent bulk measurements \cite{Zvyagin} point to
an {\it additional} transition at $T^{\ast}=9$~K. None of the
published data provide direct information on the frustration ratio
of nnn and nn interactions $\alpha =J_2/J_1$. In the present work
we show that in samples with thoroughly characterized
stoichiometry the transition at $T_c$ results in an {\it
incommensurate} helimagnetic state that persists to low
temperatures. We unambiguously determine the frustration ratio,
and propose an explanation for the precursor transition at $T_1$.
Several alternative mechanisms of magnetic ordering in what
``should'' have been a quantum-disordered gapped system are
proposed.

\begin{figure}
 \includegraphics[width=3.3in]{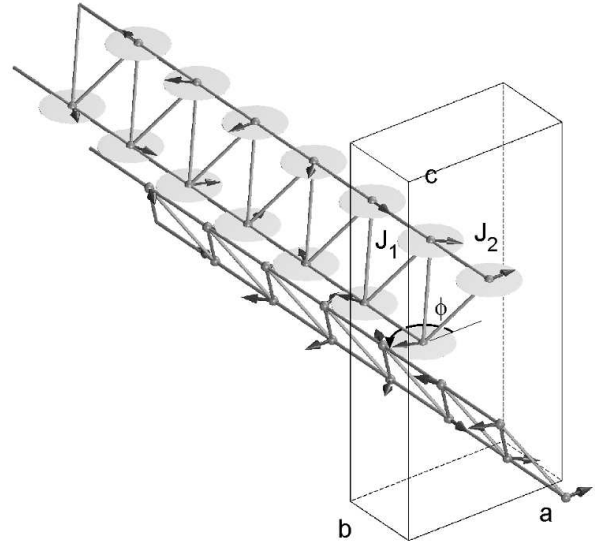}
 \caption{\label{struc} Crystallographic unit cell of \LiCuO\ showing the magnetic Cu$^{2+}$ sites (green balls)
 and the planar helimagnetic spin structure (arrows) determined in this work.}
\end{figure}

As discussed in detail in Refs.~\cite{Berger,Roessli,Zvyagin},
\LiCuO\ contains an equal number  of Cu$^{+}$ and Cu$^{2+}$ ions
in distinct non-equivalent crystallographic positions. The
magnetic Cu$^{2+}$ ions carry $S=1/2$ and form ``triangular''
two-leg ladders (Fig.~\ref{struc}), which can also be viewed as
zig-zag chains with competing nn and nnn interactions,
$J_\mathrm{nn}$ and $J_\mathrm{nnn}$, respectively. These chains
run along the $b$ axis of the orthorhombic crystal structure, and
are well separated from each other by double chains of
non-magnetic Li$^{+}$ ions and layers of non-magnetic Cu$^+$
sites. One key element of the present work was the preparation of
samples with a thorougly controlled chemical composition. Single
crystals of \LiCuO were grown in an alundum crucible in air
atmosphere using the self-flux method. The lattice parameters $a =
5.730(1)$\AA, $b = 2.8606(4)$\AA, and $c = 12.417(2)$\AA\ were
verified by powder X-rays diffraction, which also confirmed the
absence of any appreciable amounts of impurity phases. These
values are in very good agreement with those reported in
\cite{Berger}. All crystals were found to be microscopically
twinned with respect to the $[1,1,0]$ plane, so that $a\approx
2b$.

\begin{figure}
 \includegraphics[width=3.3in]{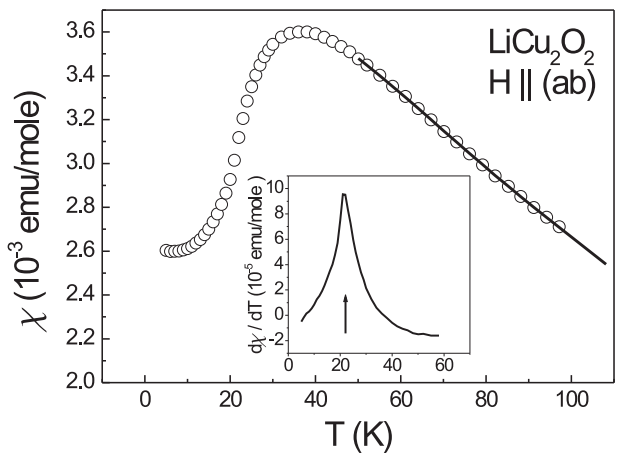}
 \caption{\label{sus} Temperature dependence of
 magnetic susceptibility  of \LiCuO\ measured in a
 magnetic field $H=100$~Oe applied parallel to the $(a,b)$
 cleavage plane (symbols). The solid line is a fit based on the frustrated
 $S=1/2$ chain model, as described in the text. Taking a numerical derivative (inset) reveals a phase transition
$T_c=22$~K (arrow).}
\end{figure}

Unexpectedly, it was found that those samples with the smallest
mosaic spread and most reproducible magnetic behavior had lower
content of Cu$^{+}$ ions than follows from the stoichiometic
formula Li$_1$+Cu$_2$+O$_2$. The density was determined to be
$\rho=5.12$~g/cm$^3$ at room temperature, which corresponds to an
actual composition Li$_{1.16}$Cu$_{1.84}$O$_{2.01}$. Chemical
disorder and a Cu-deficiency by as much as $x=16$~\% are thus
inherently present. The ``surplus'' Li$^{+}$ ions in
Li$_{1.16}$Cu$_{1.84}$O$_{2.01}$ occupy Cu$^{2+}$ sites, due to a
good match of ionic radii. A substitution of Cu$^{+}$ ions by
Li$^{+}$ ions in the \LiCuO\ structure is prevented by the
dumbbell oxygen coordination being very uncharacteristic for the
latter ion type. Charge compensation requires that the
introduction of 16\% non-magnetic Li$^{+}$ ions into the double
chains is accompanied by a transfer of 16\% of the
$S=1/2$-carrying Cu$^{2+}$ ions onto the Cu$^{+}$ inter-chain
sites.  As will become crucial for the discussion below, our {\it
disordered and non-stoichiometric}
Li$_{1.16}$Cu$_{1.84}$O$_{2.01}$ crystals (referred to as simply
``\LiCuO'' throughout the rest of the paper) have appreciable
concentrations of both {\it non-magnetic} Li$^{+}$ impurities in
the zig-zag chains, and {\it magnetic} Cu$^{2+}$ impurities
positioned in-between chains.
\begin{figure}
 \includegraphics[width=3.3in]{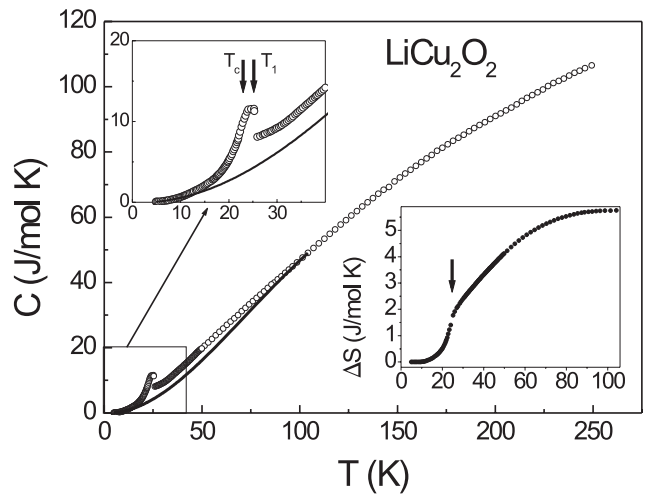}
 \caption{\label{spec} The measured temperature dependence of specific heat
 in \LiCuO (open symbols) indicates a phase transition at $T_c \approx22$~K and
 a possible precursor at $T_1 \approx25$~K (arrows). Subtracting
 the phonon contribution (solid line) allows to extract the temperature
 dependence of magnetic entropy (lower right inset). }
\end{figure}

The single crystal samples were characterized by bulk
susceptibility and specific heat measurements. $\chi(T)$ data were
taken in a commercial SQUID magnetometer in the temperature range
5--100~K and a magnetic field $H=100$~Oe applied parallel to the
$(a,b)$ cleavage plane and are shown in symbols in Fig.~\ref{sus}.
The main feature is a broad maximum at $T\approx 36$~K
characteristic of a quasi-one-dimensional magnet, that signifies
the formation of short-range correlations within the chains.
Taking the temperature derivative of the magnetic susceptibility
(Fig.~\ref{sus}, inset) reveals a sharp anomaly at $T_c=22$~K,
that we attribute to the onset of long-range magnetic order. The
high-temperature part of the $\chi(T)$ curve is expected to be
uninfluenced by this phase transition, representing the behavior
of individual zig-zag chains. The data taken above $T=50$~K were
analyzed using the high-temperature expansion formulas for the
susceptibility of the quantum $S=1/2$ frustrated-chain model
\cite{Buhler2000}. An excellent fit (solid line in Fig.~\ref{sus})
is obtained with $J_1=5.9(1)$~meV, $g=2.114(2)$, and
$\alpha=0.282(3)$. The anomaly at $T_c$ is also manifest in the
specific heat data measured using a ``Termis'' quasiadiabatic
microcalorimeter and plotted in Fig.~\ref{spec}. The peak observed
at $T\approx T_c$ is well-defined, but, as indicated by arrows in
the blowup plot, actually has a characteristic flat top that
extends between $T_c=22$~K and $T_1=24$~K, in agreement with the
results of Ref.~\cite{Roessli}. The solid line in Fig.~\ref{spec}
represents a crude estimate for the phonon contribution,
calculated under the assumption that the total magnetic entropy
released at short-range and long-range ordering in \LiCuO\ is
equal to $R\ln(2S+1) = 5.76$~J$/$mole~K. The temperature
dependence of magnetic entropy extracted in this fashion in
plotted in the lower right inset of Fig.~\ref{spec}. A large
fraction of the entropy  is released above the ordering
temperature, as expected for a low-dimensional system. The sharp
9~K anomaly reported in Ref.~\cite{Zvyagin} is totally absent in
our $\chi(T)$ and $C(T)$ data. We suspect that this feature is due
to an impurity phase, most likely Li$_2$CuO$_2$, which is known to
go through an AF transition at $9$~K \cite{Boehm}. Otherwise, the
characteristics of our samples seem to be in good agreement with
those reported by other authors
\cite{Vorotynov,Zatsepin,Fritschij98,Roessli}.

The nature of the magnetically ordered state was determined in a
single-crystal neutron diffraction experiment. Since \LiCuO\ was
prepared using a naturally occuring Li isotope mixture, the
attenuation of a neutron beam due to absorbtion by $^{6}$Li nuclei
was significant. The sample was cut parallel to the $(a,b)$
crystallographic plane to the shape of a thin plate $0.9\time
15\times 15 \times 15$~mm$^3$. The measurements were performed at
the HB1 and HB1A 3-axis spectrometers installed at the High Flux
Isotope Reactor at ORNL. For measuring integrated Bragg
intensities the instruments were used in 2-axis mode, with a
well-collimated incident neutron beam of a fixed energy
$E_i=14.7$~meV. A Pyrolitic Graphite PG(002) reflection was used
in the monochromator. A 5~cm thick PG filter was positioned in
front of the sample to eliminate higher-order beam contamination.
Bragg intensities were collected in rocking curves and corrected
for the usual Lorentz factor. Absorbtion corrections were applied
assuming a  thin-plate geometry. High-resolution measurements of
the magnetic propagation vector were performed in 3-axis mode,
with a PG(002) analyzer and $48'-40'-40'-240'$ collimators.

\begin{figure}
 \includegraphics[width=3.2in]{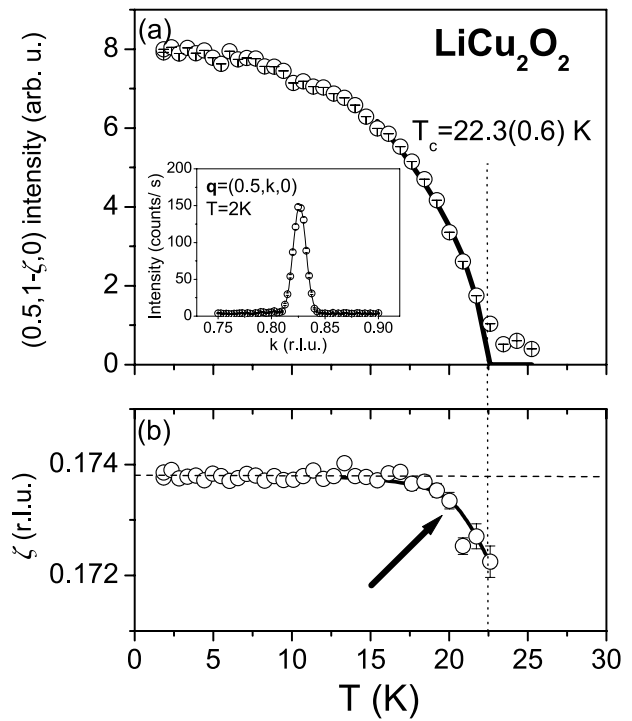}
 \caption{\label{orderp} (a) Measured temperature dependence of the $(0.5, 0.826,0)$ magnetic peak
 intensity in \LiCuO\ (symbols). The solid line is a power law fit to the data. Inset: $k$-scan across
 this reflection measured at $T=2$~K.
 (b) Measured incommensurability parameter $\zeta$
 plotted as a function of temperature. The solid line is a guide for the eye.  }
\end{figure}

The main finding of this work is that below $T_c\approx 22$~K
\LiCuO\ acquires {\it incommensurate} magnetic long range order
\footnote{ While preparing this manuscript, we became aware of an
independent unpublished NMR study by A. A. Gippius, E. N.
Morozova, A. S. Moskvin, A. V. Zalessky, A. A. Bush, M. Baenitz
and S.-L. Drechsler, that contains additional evidence of
incommensurability in \LiCuO, but provides no specific information
on the magnetic structure, frustration ratio or the role of
non-stoichiometry.}. The phase transition leads to the appearance
of new Bragg reflections that can be indexed as
$(\frac{2n+1}{2},k\pm \eta,l)$, $n,k,l$-integer,
$\zeta\approx0.174$. Such peaks were observed in both
crystallographic twins. The peak widths were found to be
resolution-limited along all three crystallographic directions at
all temperatures below $T_c$. A typical scan across the
$(0.5,0.826,0)$ reflection taken at $T=2$~K is shown in the inset
of Fig.~\ref{orderp}a. The corresponding peak intensity is plotted
as a function of temperature in Fig.~\ref{orderp}a. A simple power
law fit to the data taken above $T=15$~K yields $T_c=22.3(6)$~K
and $\beta=0.25(0.07)$. The residual intensity seen in
Fig.~\ref{orderp}a at $T>T_c$ is  due to critical scattering and
is quite broad in $q$-space.

The temperature dependence of the magnetic propagation vector was
deduced from Gaussian fits to $k$-scans across the $(0.5,0.826,0)$
peak, and is plotted in symbols in Fig.~\ref{orderp}b. Below
$T\approx 17$~K the incommensurability parameter is practically
$T$-independent and appears to have a strictly incommensurate
value $\zeta=0.1738(2)$. Interestingly, as $T_c$ is approached
from below, $\zeta$ progressively decreases as indicated by the
arrow in Fig.~\ref{orderp}b. The minimum value of $\zeta$ observed
in our experiments is about $0.172$. The temperature dependence of
$\zeta$ may be the cause for a flat-top peak in the measured
specific heat curve. In the close vicinity of $T_c$, where $\zeta$
changes rapidly, the large-period structure may undergo one or
more ``devil's staircase''-type commensuration transitions,
particularly at $T_1$ \cite{Roessli}. Intensity being the limiting
factor, we were unable to locate any well-defined magnetic Bragg
relections in the narrow temperature range $T_c<T<T_1$.

The spin arrangement in the ordered state was deduced from the
analysis of 23 non-equivalent magnetic Bragg peaks with $0 \leq h
\leq 3.5$, $0\leq k \leq 1.5$ and $0\leq l \leq 8$. Those
reflections for which the incident or scattered beams formed
angles of less that 15$^{\circ}$ with the crystalline plate were
not included in this data set, since the corresponding absorbtion
corrections are expected to deviate substantially from the
thin-plate approximation.  After examining a number of models, we
found that excellent fits to the data are obtained assuming a
planar spin-helix propagating along the zig-zag Cu$^{2+}$ chains,
with a fixed relative rotation angle $\phi=\pi (1-\zeta)$ between
consecutive spins. Any spins related by a translation along the
$c$ axis and $b$ axes are parallel and antiparallel to each other,
respectively. The magnetic structure factor for such a helical
state can be easily calculated analytically \cite{Furrer95}. After
averaging over possible $q$-domains and taking into account the
magnetic form factor of Cu$^{2+}$, an excellent fit to the
measured magnetic intensities is obtained assuming all spins to be
confined to the $(a,b)$ crystallographic plane, and adjusting only
an overall intensity scaling factor. This spin structure is
visualized in Fig.~\ref{struc}. Unfortunately, an independent
estimate of the magnitude of the ordered moment from our data was
prevented by rather severe extinction corrections of nuclear Bragg
intensities typically used for normalization.

A helimagnetic state is the typical way geometric frustration is
resolved in a classical magnet. Perhaps the first of many known
examples is that of MnO$_2$, dating back to the 50's \cite{MNO2}.
The incommesurability parameter $\zeta$ is directly related to the
frustration ratio $\alpha$ through $4\alpha=1/\cos(\pi \zeta)$. In
our case, for $\zeta=0.174$ this yields $\alpha=0.29$, in
excellent agreement with our estimates based on high-temperature
bulk susceptibility data. What makes \LiCuO\ remarkable, is that
according to existing theories for {\it quantum} $S=1/2$ spin
chains with nnn interactions, this value corresponds to a {\it
gapped disordered dimer-liquid state}, not too far from the
collinear gapless phase realized for $\alpha<0.241$ \cite{OSO}.
This would seem to agree well with the observation of a spin gap
at $T>T_c$ \cite{Zvyagin}, but is of course incompatible with
incommensurate long-range order at lower temperatures. One
possible explanation for the observed behavior are strong
inter-chain interactions, such as those that induce ordering in
the well understood Haldane-gap compound CsNiCl$_3$
\cite{CSNICL3}. Here the amplitude of transverse dispersion of the
energy gap increases with decreasing $T$, eventually driving the
gap to zero at the 3D zone-center and inducing a soft-mode phase
transition. A similar mechanism may be active in \LiCuO, and is
not incompatible with the ESR data of Ref.~\cite{Zvyagin}.
Nevertheless, it is difficult to envision the necessary strong
inter-chain superexchange pathways in the stoichiometric material.

Chemical disorder, that is {\it inherently present in our samples}
may be expected to play an important role in magnetic ordering. In
particular, the 16\% Li$^{+}$ {\it non-magnetic} defects in the
double chains will locally destroy valence bond spin-singlets and
liberate end-chain spin degrees of freedom. The free spins can
then order in three dimensions via arbitrary small inter-chain
interactions. This mechanism is known to drive magnetic ordering
in the spin-Peierls compound CuGeO$_3$ \cite{CUGEO3}, and the
Haldane-gap system PbNi$_2$V$_2$O$_8$ \cite{PBNI}. Alternatively,
long-range ordering in \LiCuO\ may be driven by the 16\% Cu$^{2+}$
impurities positioned in-between the gapped spin chains. Such {\it
magnetic} impurities carry interactions across the non-magnetic
Cu$^{+}$ layers, completing a 3D network of magnetic interactions.
Long range order occurs for an arbitrary small concentration of
impurities, and involves ordering of {\it} both the impurity spins
and the chain spins. This model in realized in
$(R_x$Y$_{1-x})_2$BaNiO$_5$ rare earth nickelates, where classical
$R^{3+}$ ions bridge the Ni-based gapped spin chains \cite{RBANO}.
To fully understand the role of impurities in the ordering of
\LiCuO\ more detailed neutron scattering experiments on
isotopically pure samples are required.

In summary, we have observed incommensurate  helimagnetism in a
$S=1/2$ zig-zag chain compound where the estimated frustration
ratio corresponds to gapped disordered phase. We speculate that
long-range magnetic ordering is aided by intrinsic impurities and
deviations from stoichiometry. This work was partially supported
by RFBR Grants 02-02-17798 and 03-02-16108. Work at ORNL was
carried out under DOE Contract No. DE-AC05-00OR22725.


\begin{thebibliography}{17}
\expandafter\ifx\csname
natexlab\endcsname\relax\def\natexlab#1{#1}\fi
\expandafter\ifx\csname bibnamefont\endcsname\relax
  \def\bibnamefont#1{#1}\fi
\expandafter\ifx\csname bibfnamefont\endcsname\relax
  \def\bibfnamefont#1{#1}\fi
\expandafter\ifx\csname citenamefont\endcsname\relax
  \def\citenamefont#1{#1}\fi
\expandafter\ifx\csname url\endcsname\relax
  \def\url#1{\texttt{#1}}\fi
\expandafter\ifx\csname
urlprefix\endcsname\relax\def\urlprefix{URL }\fi
\providecommand{\bibinfo}[2]{#2}
\providecommand{\eprint}[2][]{\url{#2}}

\bibitem[{nnn()}]{nnn}
\bibinfo{note}{C. K. Majumdar and D. K. Ghosh, J. Math. Phys. {\bf 10}, 1388
  (1969); B. S. Shastry and B. Sutherland, Phys. Rev. Lett. {\bf 47}, 964
  (1981); F.~D.~M.~Haldane, Phys. Rev. B {\bf 25}, 4925 (1982); S. R. White and
  I. Affleck, Phys. Rev. B {\bf 54}, 9862 (1996); A. A. Aligia, C. D. Batista
  and F. H. L. Essler, Phys. Rev. B {\bf 62}, 3259 (2000) and references
  therein.}

\bibitem[{\citenamefont{A.~A.~Nersesyan and Essler}(1998)}]{Nersesyan98}
\bibinfo{author}{ \bibnamefont{A.~A.~Nersesyan}}
  \bibnamefont{and} \bibinfo{author}{\bibfnamefont{F.~H.~L.}
  \bibnamefont{Essler}}, \bibinfo{journal}{Phys. Rev. Lett.}
  \textbf{\bibinfo{volume}{81}}, \bibinfo{pages}{910} (\bibinfo{year}{1998}).

\bibitem[{Vor()}]{Vorotynov}
\bibinfo{note}{A. M. Vorotynov {\it et al.}, JETP {\bf 86}, 064424 (1998); J.
  Magn. magn. Mater. {\bf 188}, 233 (1998).}

\bibitem[{\citenamefont{Fritschij et~al.}(1998)\citenamefont{Fritschij, Brom,
  and Berger}}]{Fritschij98}
\bibinfo{author}{\bibfnamefont{F.}~\bibnamefont{Fritschij}},
  \bibinfo{author}{\bibfnamefont{H.}~\bibnamefont{Brom}}, \bibnamefont{and}
  \bibinfo{author}{\bibfnamefont{R.}~\bibnamefont{Berger}},
  \bibinfo{journal}{Solid State Commun.} \textbf{\bibinfo{volume}{107}},
  \bibinfo{pages}{719} (\bibinfo{year}{1998}).

\bibitem[{Zat()}]{Zatsepin}
\bibinfo{note}{A. A Zatsepin {\it et al.}, Phys. Rev. B {\bf 57}, 4377 (1998).}

\bibitem[{Zvy()}]{Zvyagin}
\bibinfo{note}{S. Zvyagin {\it et al.}, Phys. Rev. B {\bf 66}, 064424, (2002).}

\bibitem[{Roe()}]{Roessli}
\bibinfo{note}{B. Roessli {\it et al.}, Physica B {\bf 296}, 306 (2001).}

\bibitem[{Ber()}]{Berger}
\bibinfo{note}{R. Berger, J. Less-Common Met. {\bf 169}, 33 (1991); R. Berger,
  P. Onnerund and R. Tellgren, J. Alloys Compd. {\bf 184}, 315 (1992).}

\bibitem[{\citenamefont{A.~Buhler}(2000)}]{Buhler2000}
\bibinfo{author}{\bibfnamefont{G.~S.~U.} \bibnamefont{A.~Buhler},
  \bibfnamefont{N.~Elstner}}, \bibinfo{journal}{Eur. Phys. J. B}
  \textbf{\bibinfo{volume}{16}}, \bibinfo{pages}{475} (\bibinfo{year}{2000}).

\bibitem[{Boe()}]{Boehm}
\bibinfo{note}{M. Boehm {\it et al}, Europhys. Lett. {\bf 43}, 77 (1998).}

\bibitem[{Fur()}]{Furrer95}
\bibinfo{note}{A. Furrer (Ed.), Magnetic Neutron Scattering, Part I, Chapter 5,
  World Scientific Publishing (1995).}

\bibitem[{MNO()}]{MNO2}
\bibinfo{note}{R. A. Erickson, Phys. Rev. {\bf 85}, 745 (1952); A. Yoshimori,
  J. Phys. Soc. Jpn. {\bf 14}, 807 (1959).}

\bibitem[{OSO()}]{OSO}
\bibinfo{note}{T. Tonegawa and I. Harada, J. Phys. Soc. Jpn. {\bf 56}, 2153
  (1987); K. Okamoto and K. Nomura, Phys. Lett. A {\bf 169}, 433 (1992); R. D.
  Somma and A. A. Aligia, Phys. Rev. B {\bf 64}, 024410 (2001).}

\bibitem[{CSN()}]{CSNICL3}
\bibinfo{note}{W. J. L. Buyers {\it et al.}, Phys. Rev. lett. {\bf 56}, 371
  (1986); R. M. Morra {\it et al.}, Phys. Rev. B {\bf 38}, 543 (1988).}

\bibitem[{CUG()}]{CUGEO3}
\bibinfo{note}{M. Hase {\it et al.}, Phys. Rev. Lett. {\bf 71}, 4059 (1993); T.
  Masuda {\it et al.}, Phys. Rev. Lett. {\bf 80}, 4566 (1998).}

\bibitem[{PBN()}]{PBNI}
\bibinfo{note}{Y. Uchiyama {\it et al.}, Phys. Rev. Lett {\bf 83}, 632 (1999);
  Physica B {\bf 284-288}, 1641 (2000); A. Zheludev {\it et al.}, Phys. Rev. B
  {\bf 64}, 134415(2001); A. I. Smirnov {\it et al.}, Phys. Rev B {\bf 65},
  174422 (2002).}

\bibitem[{RBA()}]{RBANO}
\bibinfo{note}{A. Zheludev {\it et al.}, Phys. Rev. Lett. {\bf 80}, 3630
  (1998); S. Maslov and A. Zheludev, Phys. Rev. Lett. {\bf 80}, 5786 (1998); A.
  Zheludev {\it et al.} J. Phys.: Condens. Matter {\bf 13}, R525 (2001).}

\end{thebibliography}

\end{document}